\documentclass[fleqn,10pt]{wlscirep}
\usepackage[utf8]{inputenc}
\usepackage[T1]{fontenc}
\usepackage{subfigure}
\usepackage{float}
\usepackage{braket}
\usepackage{amssymb}
\usepackage{amsfonts}
\usepackage{amsmath}
\usepackage{physics}
\title{Long Distance Entanglement and High-Dimensional Quantum Teleportation in The Fermi-Hubbard Model}

\author[1,+]{Sanaa Abaach}
\author[2,*,+]{Zakaria Mzaouali}
\author[1,+]{Morad El Baz}
\affil[1]{ESMaR, Faculty of Sciences, Mohammed V University in Rabat, Morocco.}
\affil[2]{Institute of Theoretical and Applied Informatics, Polish Academy of Sciences, Bałtycka 5, 44-100 Gliwice, Poland.}

\affil[*]{zmzaouali@iitis.pl}

\affil[+]{these authors contributed equally to this work}

\begin{abstract}
The long distance entanglement in finite size open Fermi-Hubbard chains, together with the end-to-end quantum teleportation are investigated. We show the peculiarity of the ground state of the Fermi-Hubbard model to support maximum long distance entanglement, which allows it to operate as a quantum resource for high fidelity long distance quantum teleportation. We determine the physical properties and conditions for creating scalable long distance entanglement and analyze its stability under the effect of the Coulomb interaction and the hopping amplitude. Furthermore, we show that the choice of the measurement basis in the protocol can drastically affect the fidelity of quantum teleportation and we argue that perfect information transfer can be attained by choosing an adequate basis reflecting the salient properties of the quantum channel, i.e. Hubbard projective measurements.
\end{abstract}
\begin{document}

\flushbottom
\maketitle
%
%
\thispagestyle{empty}

\section{Introduction}

The second quantum revolution is driven by the revolutionary ideas of exploiting the inherent quantum properties of atomic systems in order to achieve a quantum advantage over classical methods in the manipulation of information~\cite{nielsen2002quantum, lars2018second}. One of the facets of quantum information processing is quantum teleportation, which is a protocol offering the possibility of transferring an unknown quantum state using pre-existing entanglement and a classical information channel~\cite{EPR_teleportation}. The introduction of the protocol in 1993 by Bennett \textit{et al}, and its experimental realization in 1997 by the group led by Zeilinger~\cite{zeilinger1997} shifted the concept of teleportation from being fictional to a physical reality~\cite{popescu1998exp}.

Quantum teleportation rely primarily on the entanglement shared between the sending and receiving party through the quantum channel~\cite{miranowicz2003introduction}. Therefore, the creation and distribution of entanglement in physical platforms is crucial in the success and implementation of quantum teleportation protocols~\cite{Campos_Venuti}. In this context, a variety of one-dimensional quantum spin chains are known for their entangled ground states and have been intensively investigated as faithful architectures for quantum information processing~\cite{QIProcessing1, QIProcessing2, QIProcessing3, QIProcessing4, amico2008entanglement, abaach2021pairwise, Mzaouali2018, Mzaouali2019} and notably as reliable quantum channels for teleportation protocols~\cite{nikolopoulos, APOLLARO, Quantum_transfer, Apollaro_2022, yousefjani2021parallel, hermes2020dimensionality, pouyandeh2015quantum, lorenzo2015transfer, yang2015transfer, bayat2014arbitrary, apollaro2015many}. Nevertheless, in most one-dimensional quantum spin chain systems with short-range interactions, the entanglement vanishes for distances larger than two neighboring sites~\cite{range1, range2, range3}, which makes them inconceivable platforms for long distance teleportation.

Accordingly, efforts have been devoted to produce some mechanisms able to create sizable entanglement between distant but not necessarily directly interacting constituents. An early initiative was the introduction of the concept of localizable entanglement, which defines the concentrated entanglement on an arbitrary distant pair by implementing optimal local measurements onto the rest of the system~\cite{Campos_Venuti}. Such kind of entanglement is defined as long distance entanglement (LDE).

A promising candidate for creating LDE is the gapped one-dimensional antiferromagnetic spin chain. In this regard, Venuti \textit{et al}~\cite{Campos_Venuti} have proposed a scheme for creating LDE in spin-$\frac{1}{2}$ and spin-1 chains, showing that this property appears only for given values of a specific microscopic parameter, which doesn't coincide with known quantum critical points. Later, schemes have been suggested for generating LDE in $XX$ spin chains~\cite{Long_teleportation_XX, Generation_of_Long-Distance}, in many-body atomic and optical systems~\cite{Giampaolo_2010}, in Motzkin and Fredkin spin chains~\cite{Motzkin_Fredkin}, and in antiferromagnetic $XXZ$ spin chain with alternating interactions~\cite{HU2022} as well. Additionally, it has been shown that robust temporally shaped control pulses for producing LDE can be derived in disordered spin chains~\cite{Cui_2015} and that LDE can be enhanced for spin chains with dissipative processes through global measurements~\cite{Rafiee_2018}. Recently, an experimental implementation of LDE has been realised between unpaired spins in antiferromagnetic spin-$\frac{1}{2}$ chains in a bulk material~\cite{sahling2015experimental}. As a consequence of the appealing LDE generation property, it has been recently demonstrated that LDE allows for robust qubit teleportation and state transfer with high fidelity, across sufficiently long distances in finite size spin chains~\cite{Qubit_Teleportation, Long_teleportation_XX}. 

So far, investigating the generation of LDE has been limited to qubit spin chains. However, increasing the dimensionality and the complexity of the system has been shown to enhance the capacity of the quantum communication channel~\cite{qutrits, highD2}, and improves the robustness against eavesdropping attacks~\cite{highD3, highD4}. Additionally, high-dimensional entangled states can be used for quantum state transfer of ever-increasing complexity~\cite{highD5}. Similarly, the majority of quantum teleportation experiments, were limited to two-dimensional subspaces (qubits), including quantum dot spin qubits~\cite{QD1}. Recently, quantum teleportation has been achieved in high-dimensional quantum photonic systems~\cite{photonic, slaoui}. An alternative for photonic platforms, which causes the propagation losses of light, are quantum dots as they are the most scalable and time coherent architectures dedicated for quantum simulation and implementing quantum information tasks in the form of communication and computing~\cite{nichol2022quantum}. Recently, it has been shown that quantum dot systems described by the ground state of the Hubbard model are a promising entanglement resource for performing quantum teleportation of four-dimensional states~\cite{Long_rangeQD}.

The main goal of this paper is to analyze long distance quantum teleportation in quantum dots described by the ground state of the one-dimensional Fermi-Hubbard model. We start in Section~\ref{sec: section2} by introducing the general form of the Fermi-Hubbard Hamiltonian, and presenting the generalized standard teleportation protocol with an arbitrary mixed state resource in higher dimensions, as well as the fidelity of quantum teleportation. Moreover, we define the lower bound of concurrence for measuring the end-to-end entanglement in bipartite high dimensional states. The aforementioned framework allows us in Section.~\ref{sec: section3} to discuss two schemes for creating end-to-end entanglement in an open Fermi-Hubbard chain, one by implementing bonds of alternating strengths defined in~\cite{Campos_Venuti} and the other by alternating hopping amplitudes. Additionally, we show the important interplay between the choice of the measurement basis in the protocol and achieving perfect quantum teleportation with unit fidelity. Finally, in Section.~\ref{sec: section4} we summarize our core results and conclusions.
\section{\label{sec: section2}The model and the teleportation protocol}

\subsection{The Fermi-Hubbard model}
The Fermi-Hubbard (FH) model describes moving fermions with spin in a lattice~\cite{ref21}. In one-dimensional settings, it is given by
\begin{equation}
H = -t \sum_{i,\sigma} \left( c_{i,\sigma}^{\dag} c_{i+1,\sigma}+ c_{i+1,\sigma}^{\dag} c_{i,\sigma} \right) + u \sum_{i} n_ {i,\uparrow} n_{i,\downarrow},    
\label{ham}
\end{equation}
where $c_{i,\sigma}^{\dag}$ and $c_{i,\sigma}$ are, respectively, the creation and annihilation operators that describe the tunneling of electrons between the neighboring sites. $t$ is the hopping amplitude and $\sigma\!=\!\{ \uparrow,\downarrow \}$ indicates spin-up or spin down electron, whereas $u$ is the on-site electron-electron Coulomb interaction. We assume that only the $s$-orbital is allowed to the electrons in each site, so that, each site is able to hold up to two electrons with opposite spins as stated by the Pauli exclusion principle. Thereby, electrons have four possibilities in occupying a single site: $\ket{0}$, $\ket{\uparrow}$, $\ket{\downarrow}$ and $\ket{\uparrow\downarrow}$.
When the Coulomb interaction $u$ is strong enough, the tunneling of electrons between the sites $t$ is blocked leading to the quantum confinement effect in the FH system. This physical picture is analogous to the formation of potential barriers between the sites that prohibits electrons to tunnel outside. Experimentally, the creation of such barriers is made by modulating potentials, using gate electrodes, in order to control the tunneling of electrons between the sites that are simulated using semiconductor quantum dots~\cite{ref9}. 

The Fermi-Hubbard Hamiltonian, Eq.~\eqref{ham}, is a prototype model to describe and investigate the properties of quantum systems, such as: the metal-insulator transition, ferromagnetism, ferrimagnetism, and antiferromagnetism. As well as, superconductivity and Tomonaga-Luttinger liquid~\cite{essler2005one}. Furthermore, the intersection between quantum information theory and condensed matter physics has been a subject of interest in fermionic models. In particular, it has been shown that entanglement plays a role in the identification of the phases of matter present in the Hubbard model~\cite{anfossi2007entanglement, gu2004entanglement, kleftogiannis2019exact, canella2019superfluid, harir2019quantum, spalding2019prb, zawadzki2020work}. At the experimental level, quantum simulation of the Hubbard model using ultra cold atoms in optical lattices have been successful in observing and validating the theoretical results of strongly correlated Fermi gases~\cite{2018_CRP_Tarruell}. In the following, we will consider the dimensionless quantity $U\!=\!u/t$ as the main parameter in the model, Eq.~\eqref{ham}.

\subsection{Quantum Teleportation with $\mathbf{d}$-dimensional channels}
The standard quantum teleportation protocol with Bell states resources is an example of a noiseless channel. It allows the transfer of information from a sender $A$ to a receiver $B$ using Bell measurements and Pauli rotations. In general, quantum teleportation uses entangled mixed states as a resource, which renders the protocol equivalent to information transfer via a noisy channel~\cite{bowen2001teleportation}. To teleport information encoded in a $d-$dimensional unknown state $\ket{\Psi}\!=\!\sum_{j=0}^{d-1} \alpha_{j} \ket{j}$, where $\alpha_{i} \in \mathbb{C}$, the teleportation protocol have to operate using an entangled two qudit ($d\!\times\!d$) state as a state resource. Hence, the teleportation protocol is constructed and formulated by means of the maximally entangled state $\ket{\psi^{+}}= \frac{1}{\sqrt{d}}\sum_{j}\ket{j}\ket{j}$, and the set of unitary generators $U^{nm}=\sum_{k} e^{2\pi i k n /d} \ket{k}\bra{k\oplus m}$ that act on the first subsystem, where $\oplus $ denotes the addition modulo $d$. Interestingly, it has been shown that the output state is equivalent to the state produced by a depolarizing channel~\cite{bowen2001teleportation}, given by
\begin{equation}
    \epsilon( \varrho )= \sum_{nm} \Tr[E^{nm}\chi] U^{n(-m)}\varrho (U^{n(-m)})^\dag,
    \label{output_state}
\end{equation}
where $\chi$ denotes the quantum state resource or the shared entangled mixed state between the sender and the receiver in a teleportation protocol, and $E^{nm}$ are the set of maximally entangled Bell state projectors given by $E^{nm}\!=\!U^{nm}\ket{\psi^{+}}\bra{\psi^{+}}{(U^{nm})^\dag}$, with $n, m\!=\!0, 1,\dots, d-1$. 

The standard quantum teleportation channel being noisy hinders the perfect information transfer using a general $(d\!\cross\!d)$ bipartite state as a resource, instead of $\ket{\psi^{+}}$. Accordingly, measuring how well the output state $  \epsilon( \varrho) $ and the input state $\varrho$ are similar in a quantum teleportation scheme is quantified through the concept of the fidelity~\cite{fidelity_mixed}. It is defined as

\begin{equation}
    \mathcal{F}= \Tr(\varrho \epsilon(\varrho)).
    \label{fidelity}
\end{equation}
For orthogonal states the fidelity is zero, while it reaches a unit value for identical states. Correspondingly, the efficiency and the quality of a quantum channel in teleporting an unknown state is presented in terms of the average fidelity $\overline{\mathcal{F}}$ over all possible input states, which can be written as
\begin{equation}
    \overline{\mathcal{F}}= \frac{d}{d+1}f + \frac{1}{d+1},
\end{equation}
where
\begin{equation}
    f= \bra{\psi^+}\chi\ket{\psi^+},
     \label{fraction}
\end{equation}
and $\chi$ \ is the quantum state resource. $\overline{\mathcal{F}}(\epsilon)$ is proven to be the maximal achievable teleportation fidelity in the standard teleportation protocol, where $f$ is the fully entangled fraction \cite{fraction_1, fraction_2,fraction_3}. Indeed, one requires the average fidelity to be larger than $\frac{2}{d+1}$ to outperform the purely classical communication protocols. Hence, for a two-qudit state $\chi$, as a resource to be useful for quantum teleportation, the fully entangled fraction $f$ needs to satisfy $f\!>\!\frac{1}{d}$.
\begin{figure}[t!]
        \subfigure[\label{fig1a}]{%
       \includegraphics[width=0.5\textwidth]{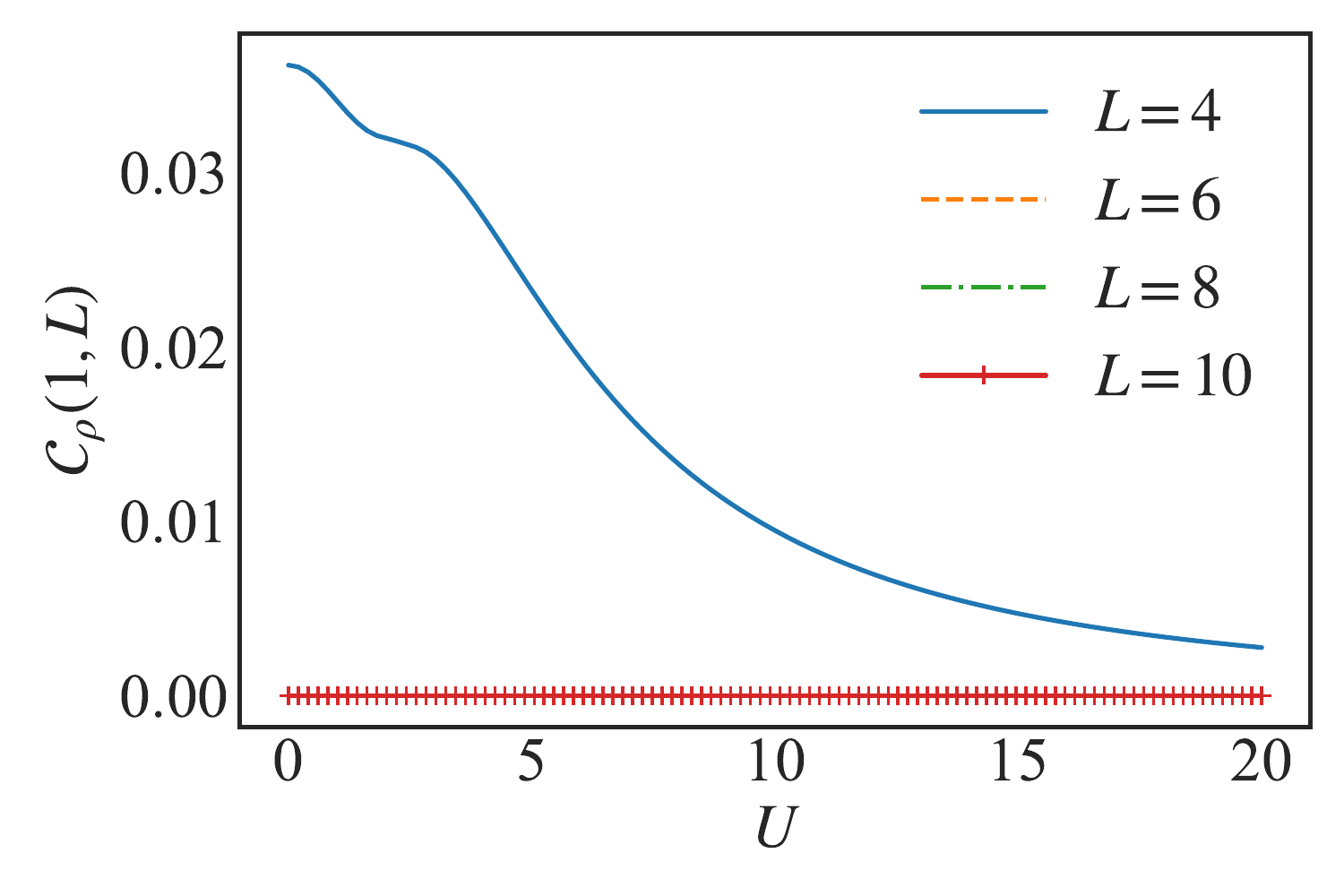}
       }%
     \subfigure[\label{fig1b}]{%
       \includegraphics[width=0.5\textwidth]{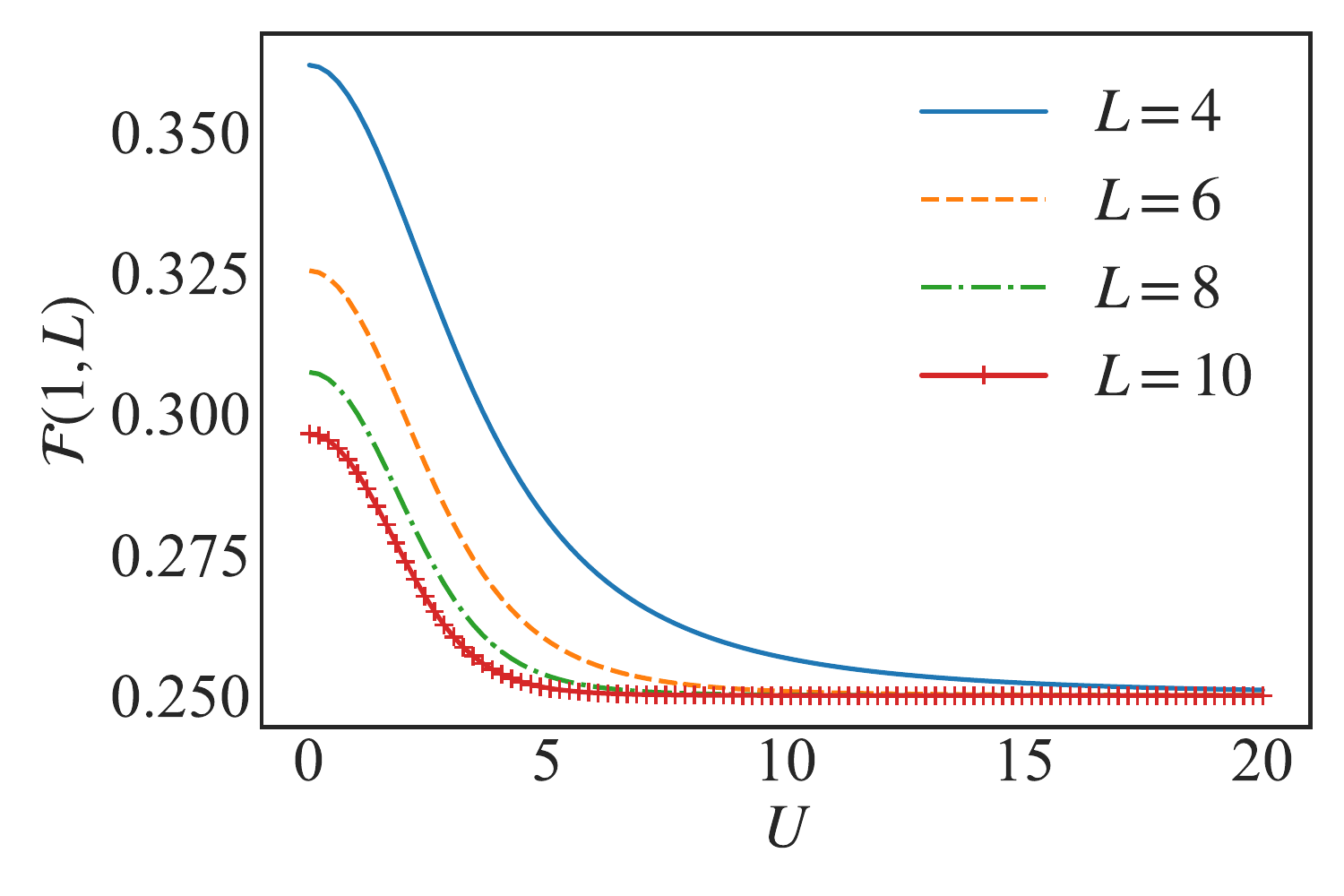}
       }
     \caption{(a) The end-to-end concurrence, Eq.~\eqref{lbc}, and (b) the fidelity, Eq.~\eqref{fidelity}, versus $U$ in the Fermi-Hubbard model, Eq.~\eqref{ham}, for several chain sizes $L$.}
     \label{fig1}
\end{figure}
\begin{figure*}[t!]
  \centering
  \subfigure[]{\includegraphics[width=0.33\textwidth]{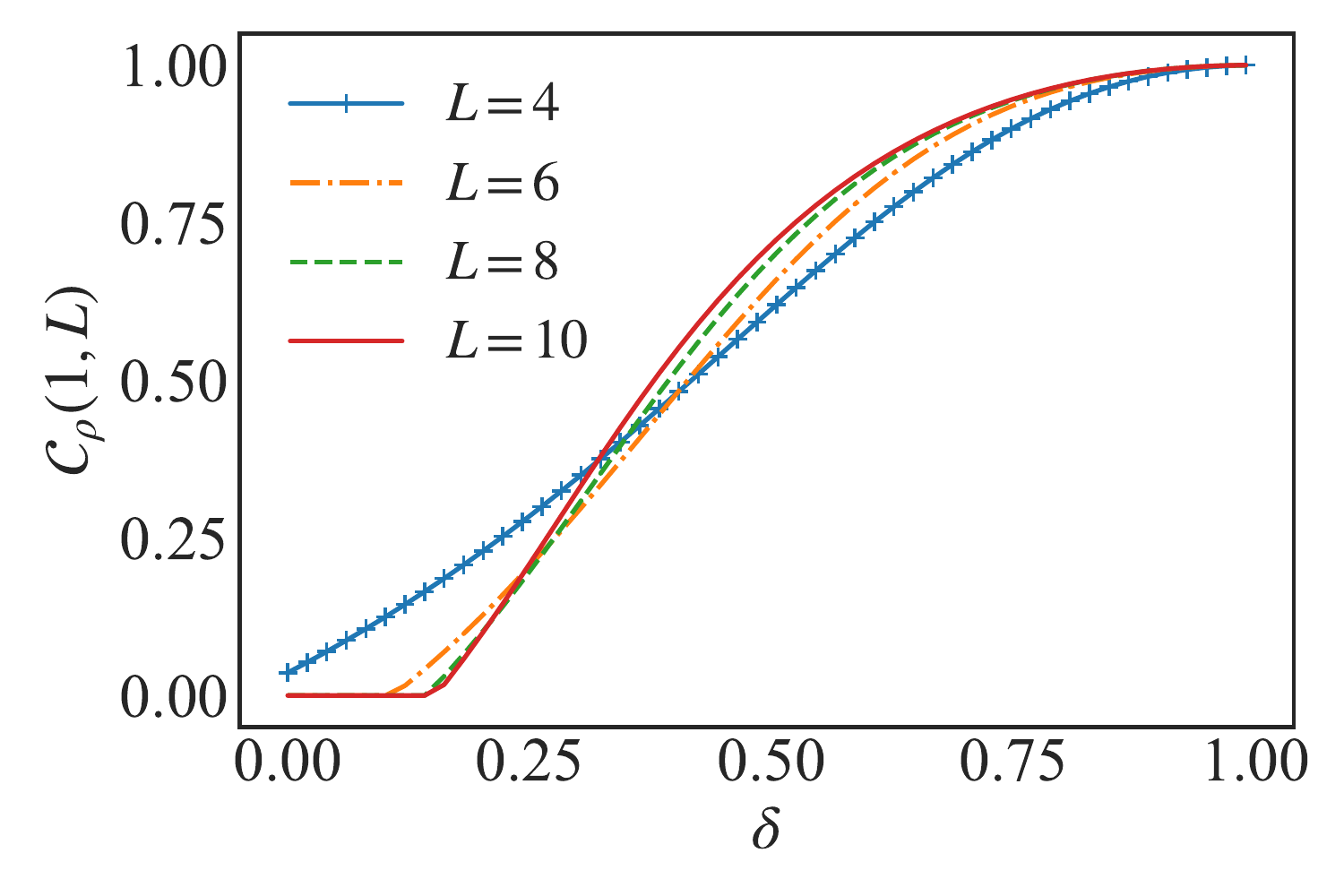}
  \label{fig2a}}%
  \subfigure[]{\includegraphics[width=0.33\textwidth]{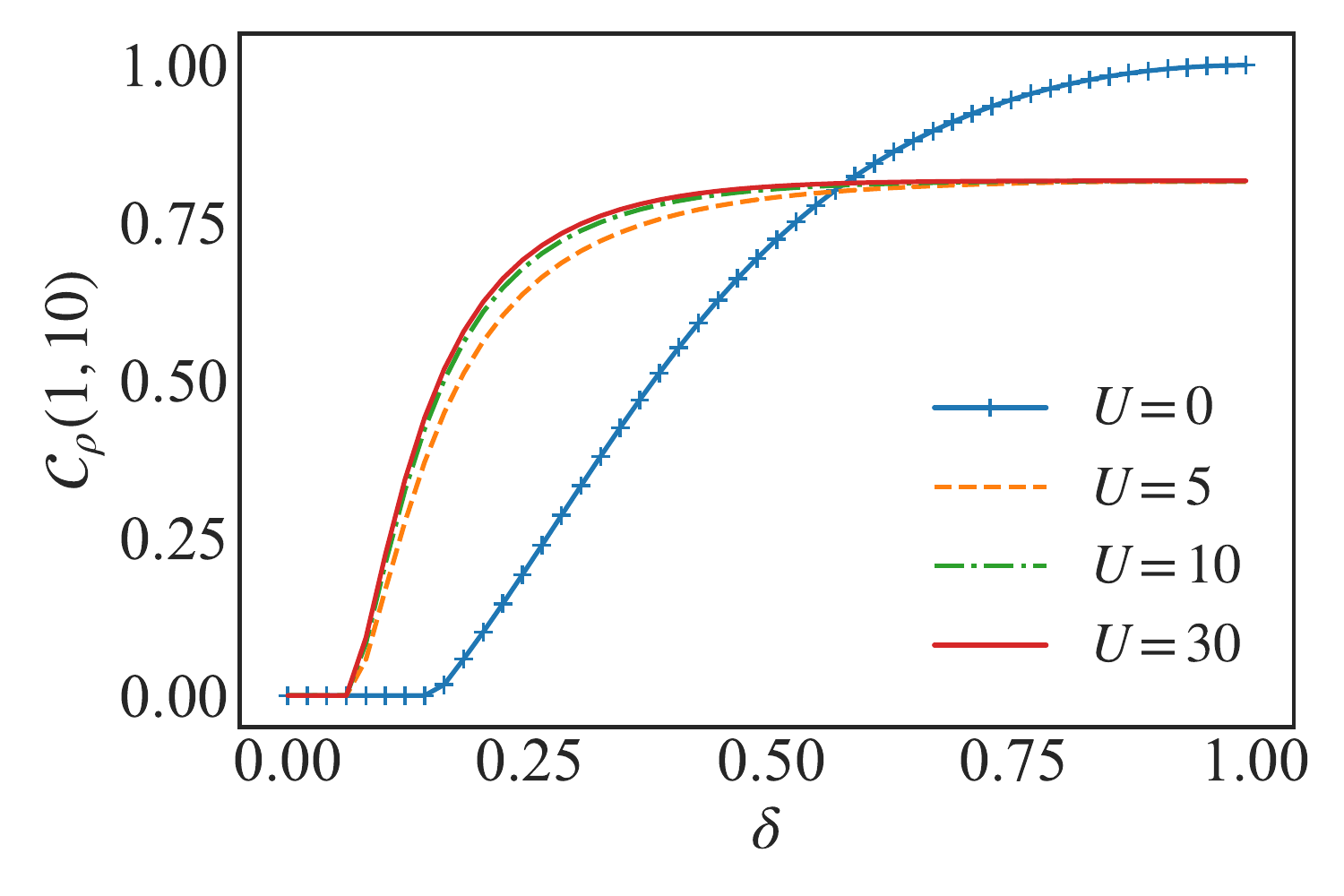}
  \label{fig2b}}%
  \subfigure[]{\includegraphics[width=0.33\textwidth]{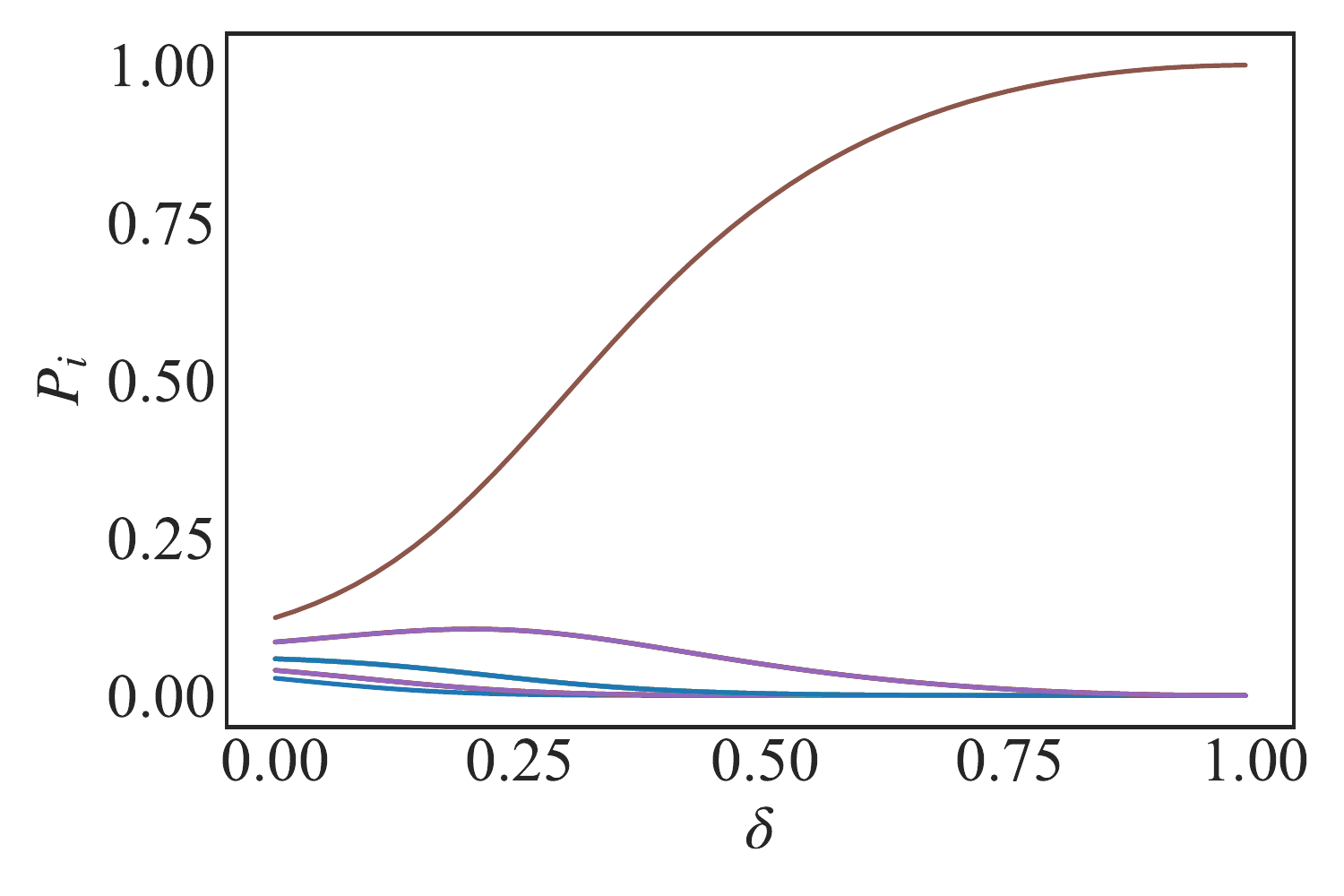}
  \label{fig2c}}
  \caption{(a)+(b) The end-to-end concurrence, Eq.~\eqref{lbc}, with respect to $\delta$ for different chain sizes $L$, and for various values of $U$ with fixed $L$, respectively. (c) Evolution of the occupation probabilities $P_i$ of the local half filled state with respect to $\delta$.}
  \label{fig:img1}
\end{figure*}

The key ingredient for the success of the teleportation protocol is the entanglement  shared between the sender and receiver, that is actually the quantum state resource $\chi$. Therefore, quantifying the entanglement in quantum systems is crucial in order to evaluate and improve the fidelity of the quantum teleportation protocol. Accessing the amount of entanglement in many-body quantum systems is a challenging task and exact formulas are known only for some configurations, e.g. qubit-qubit, and qubit-qutrit systems~\cite{horodecki2009}.  In~\cite{reff6} the authors introduced an analytical lower bound of concurrence as an effective evaluation of entanglement for arbitrary bipartite quantum states by decomposing the joint Hilbert space into many $2\!\otimes\!2 $ dimensional subspaces and without any optimization procedure. This lower bound was recently extended to arbitrary $N$-partite states \cite{reff7,ref13} that provides better estimates of the entanglement in some states comparted to the usual separability criteria. 

For an arbitrary mixed state in $d\!\times\!d$ dimension, the concurrence $C(\rho_{ij})$ \cite{reff6} satisfies,
\begin{equation}
  \tau_{2}( \rho_{ij})=\frac{d}{2(d-1)}\sum_{\alpha}^{\frac{d(d-1)}{2}}\sum_{\beta}^{\frac{d(d-1)}{2}} C_{\alpha\beta}^{2}\le C^{2}(\rho_{ij}),
  \label{pairlow}
\end{equation}
where,
\begin{equation}
C_{\alpha\beta}=max\{0,\lambda_{\alpha\beta}^{(1)}-\lambda_{\alpha\beta}^{(2)}-\lambda_{\alpha\beta}^{(3)}-\lambda_{\alpha\beta}^{(4)} \}.
\label{lbc}
\end{equation}
In our case, $ \rho_{ij}$ is the pairwise density matrix of the ground state of the Fermi-Hubbard model, Eq.~\eqref{ham}, with $(i\!<\!j)$. $\lambda_{\alpha\beta}^{(\hbox{a})} $ are the square roots of the non-zero eigenvalues of the non-Hermitian matrix $ \rho_{ij}\tilde{\rho}_{(ij)\alpha\beta} $ such that 
$ \lambda_{\alpha\beta}^{(\hbox{a})}\!>\!\lambda_{\alpha\beta}^{(\hbox{a}+1)} $ for $ 1\le \hbox{a}\le 3$ and
\begin{equation}
   \tilde{\rho}_{(ij)\alpha\beta}=(G_{\alpha}\otimes G_{\beta})\rho_{ij}^{*}(G_{\alpha}\otimes G_{\beta}).
\end{equation}
Here, $G_{\alpha}$ is the $\alpha^{\text{\tiny th}}$ element of the group $SO(d)$.

In the following, using QuSpin~\cite{ref31,ref32} and QuTiP~\cite{qutip1,qutip2}, we calculate by numerical diagonalization the ground state of the Fermi-Hubbard model, Eq.~\eqref{ham}, in order to exploit it as a resource for teleporting an unknown four-dimensional state $\ket{\Psi}\!=\! \frac{1}{N}(\alpha_{0} \ket{0} + \alpha_{1} \ket{1} + \alpha_{2} \ket{2} + \alpha_{3} \ket{3}$), where $\alpha_{i} \in \mathbb{C}$ and $N$ is the normalization constant. Figure 1 in the supplementary materials shows the fidelity $\mathcal{F}$, Eq.~\eqref{fidelity}, as a function of the coefficient $\alpha_{0}$, where the state to be teleported is $\ket{\Psi}= \frac{1}{\sqrt{\alpha_{0}^2 +3}}(\alpha_{0} \ket{0}+ \ket{1}+\ket{2}+\ket{3})$, $\forall \alpha_0 \in \mathbb{R}$. We report that (c.f Figure~1 in supplementary material) $\mathcal{F}$ starts with a small value at $\alpha_0=0$. Then as $\alpha_0$ tends to $1$, $\mathcal{F}$ attains a maximum value and decays asymptotically as $\alpha_0$ continue to increase ($\alpha_0>1$). Such behavior indicates that the four dimensional teleportation protocol performance depends essentially on the type of the state to be teleported. Indeed, when $\alpha_0=0$ or $\alpha_0>1$, the state $\ket{\Psi}$ could be considered respectively as a qutrit state $\ket{\Psi}=\frac{1}{\sqrt{3}}(\ket{1}+\ket{2}+\ket{3})$ or a qubit state $ \ket{\Psi} \simeq \ket{0}$. For both types the quantum teleportation fidelity (c.f Figure 1 in supplementary material) reveals small values in contrast to the quadrit state type, with $\alpha_0 =1$, where a maximum value is achieved. This implies that the efficient performance of the four dimensional teleportation channel, Eq.~\eqref{output_state} necessitates a quadrit input state with $\alpha_i\ne 0$. For this reason, in order to study the behavior of the channel's fidelity, Eq.~\eqref{fidelity}, under the effect of the Coulomb interaction $u$ and the hopping amplitude $t$, we consider in the following $\alpha_{i}= \frac{1}{2}$, with $i=0, 1, 2, 3$.

Fig.~\ref{fig1a} depicts the end-to-end entanglement quantified using the lower bound concurrence, Eq.~\eqref{lbc}, in the ground state of the Fermi-Hubbard chain, Eq.~\eqref{ham}, for several system sizes. We report that long distance entanglement manifests weakly only for the chain size of $L=4$, which reflects in the behaviour of the end-to-end fidelity of the quantum teleportation represented in Fig.~\ref{fig1b} for the same model parameters. The fidelity shows no quantum advantage as it is below the value of $\frac{2}{5}\!=\!0.4$, which is the upper limit for the classical threshold. In the following, we discuss how to generate long distance entanglement in the Fermi-Hubbard model by manipulating the nature of interactions between the sites, in order to demonstrate a quantum advantage in the quantum teleportation fidelity.

\section{\label{sec: section3}Long distance entanglement and fidelity enhancement} 

Consider a chain consisting of $L$ sites. In order to create localizable entanglement between the ends of the chain, the interaction between the block of spins, separating the ends of the chains, must be weak with the first and last sites. The localizable entanglement offer a solid framework to exploit quantum many-body systems as quantum channels for transferring information between two distant parties~\cite{Campos_Venuti}. 
\subsection{The Hubbard model with alternating bonds}

One way to achieve localizable entanglement in fermionic systems is to model the interaction between the spins with bonds of alternating strengths $(1-\delta)$ (weak bond) and $(1+\delta)$ (strong bond) with $0\le \delta \le 1$. Accordingly, the Fermi-Hubbard model, Eq.~\eqref{ham}, transforms onto

\begin{equation}
 \label{ham_delta}
    H= -\sum_{i,\sigma}(1+(-1)^{i}\delta)\left( c_{i,\sigma}^{\dag}c_{i+1,\sigma}+c_{i+1,\sigma}^{\dag} c_{i,\sigma} \right)+U \sum_{i} n_{i,\uparrow}n_{i,\downarrow}.
\end{equation}
Choosing $L$ to be even and $0\le \delta \le 1$ ensures that the spins at the end of the chain interact with a weak bond of strength $(1-\delta)$ with their respective neighbors.

The Hubbard model supports a sizable end-to-end entanglement for specific values of the microscopic parameter $\delta$. The end-to-end concurrence, Eq.~\eqref{lbc}, in the ground state of the Fermi-Hubbard model, Eq.~\eqref{ham_delta}, is plotted in Fig.~\ref{fig2a} as a function of the parameter $\delta$ for various system sizes $L$ and $U=0$. The numerical data shows the creation of long distance entanglement as the parameter $\delta$ is increased, and reaches, asymptotically, the unit value as $\delta$ tends to 1. In conjunction with that, the threshold $\delta_{T}$ indicating the birth of long distance entanglement grows with the system size $L$. This is related essentially to the size effect. Indeed, as the system size increases, the amount of the end-to-end entanglement decreases, and with the increase of $\delta$, this allows to generate new quantum correlations that will be added to the preexisting ones, when $\delta=0$, which allows thus for the early appearance of the threshold as the system size deceases.

For instance, for $L=4$ the long distance entanglement appears from $\delta=0$, because already the end-to-end concurrence associated with the pair $\rho_{1,4}$ in the ground state of the Fermi-Hubbard model Eq.~\eqref{ham} exhibits a non-vanishing value, as sketched in Fig.~\ref{fig1a}. In this case, the increase of $\delta$ allows for an immediate increase of the long distance entanglement. However, increasing the system size to $L=6$ for example, it is clear from Fig.~\ref{fig1a} that the end-to-end concurrence vanishes for the pair $\rho_{1,6}$, and in this case a threshold $\delta_{T}$ manifests around $0.2$ for $L\!\geq\!6$, to create long distance entanglement.~(c.f. Fig.~\ref{fig2a})

Nevertheless, a rapid assent to the asymptotic value of long distance entanglement is clear in Fig.~\ref{fig2a} as the size of the chain $L$ increases. As a matter of fact, as $L$ grows, the pairwise entanglement at the borders $\rho_{1,2}$ and $\rho_{L-1,L}$ declines~\cite{Long_rangeQD}, and since the role of $\delta$ is to locate great amount of entanglement between the end sites ($\rho_{1}$ and $\rho_{L}$ ) by excluding entanglement inside the pairs ($\rho_{1,2}$ and $\rho_{L-1,L}$) at the borders, the rate of degradation of the entanglement inside $\rho_{1,2}$ and $\rho_{L-1,L}$ becomes faster as $L$ grows with increasing $\delta$, which accelerate the creation of long distance entanglement between the ends of the chain. 

Fig.~\ref{fig2b} shows the effect of the Coulomb interaction $U$ on the end-to-end concurrence, Eq.~\eqref{lbc}, of the ground state of the Fermi-Hubbard model, Eq.~\eqref{ham_delta}, with respect to the parameter $\delta$, for $L\!=\!10$. When $U\!=\!0$, the long distance entanglement grows slowly but reaches  asymptotically the unit value as $\delta$ tends to one. However, for non zero values of $U$, the entanglement grows rapidly up to the asymptotic value of 0.8. Moreover, the threshold $\delta_{T}$ marking the birth of long distance entanglement, diminishes for $U\!>\!0$. This behaviour can be explained by the fact that the state of the pairs at the ends $\rho_{1,L}$ is described by the mixture $\sum_{i}P_{i}\ket{\psi}\bra{\psi}$ which is generally dominated by the local half filled state (LHFS) \cite{Long_rangeQD} associated with the probability $P_{LHFS}$. For $U\!=\!0$ the LHFS is given by the state $ \ket{\psi}_{U=0}\!=\!\frac{1}{2}( \ket{\uparrow, \downarrow} +  \ket{ \downarrow , \uparrow} + \ket{\uparrow \downarrow , 0} + \ket{0, \uparrow \downarrow})$, which indicates the free motion of electrons between the sites. Nevertheless, the increase of $\delta$ reduces the effect of the hopping inside the pairs at the borders, which in turn reduces the quantum correlations in the form of entanglement, due to the probability $P_{LHFS}$ becoming insignificant. In contrast, according to the monogamy principle (Ref), the end-to-end entanglement grows since, in this case, $P_{LHFS}$ associated to the maximally entangled pure state $\psi_{LHFS}$ increases and reaches the maximum value 1 at $\delta\!=\!1$, as shown in Fig.~\ref{fig2c}. This explains the maximum attainable value 1 of the end-to-end entanglement at $\delta\!=\!1$ when $U\!=\!0$. 

Switching on the interaction $U$ ($u/t\!>\!0 $), increases the in-site repulsion interaction which makes the electrons avoid the state of double occupancy. In this case, the local half filled state evolves with $\delta$ into the antiferromagnetic state $\ket{\psi}_{U>0}\!=\!\frac{1}{\sqrt{2}}( \ket{\uparrow, \downarrow} +  \ket{ \downarrow , \uparrow} )$ at $\delta\!=\!1$. Simultaneously, $P_{LHFS}$ associated to this state evolves into the maximum unit value at $\delta=1$. In such circumstances, $\rho_{1,L}$ is well described by a maximally entangled pure state. However, the quantum correlations contained in this state are less than those contained in $\ket{\psi}_{U=0}$, and for this reason the concurrence, Eq.~\eqref{lbc}, saturates at 0.8 instead of 1 when $U\!=\!0$.

We turn our attention to the behaviour of the fidelity of quantum teleportation when the channel is described by the ground state of the Fermi-Hubbard model, Eq.~\eqref{ham_delta}. In this regard, we have plotted the fidelity $\mathcal{F}$ as a function of the parameter $\delta$ for different values of the coefficients $\alpha_i$ (c.f Figure 3 in supplementary material) where it is shown that the quantum teleportation fidelity is independent of the state to be teleported. This indeed conforms to the results found in \cite{Campos_Venuti}, where the teleportation fidelity has been derived analytically only in terms of the parameter $\delta$. Accordingly, any input state (the state to be teleported) can be used in the quantum teleportation process. For instance, we choose a four dimensional state with $\alpha_i =\frac{1}{2}$ where $i=0,1,2,3$.

Fig.~\ref{fig3a} shows the behavior of the fidelity as a function of $\delta$ for different values of the system size $L$, at $U\!=\!0$. We see that for weak values of $\delta$, the smaller the system size, the faster the fidelity rises up above the classical threshold $\mathcal{F}\!=\!\frac{2}{5}$ (dotted line). Nevertheless, as soon as the fidelity rise above the classical threshold, it grows rapidly with $\delta$ as the system size increases, reaching the saturation value 0.5 in the limiting case $\delta\!\to\!1$. In the following, we will show how to bypass this saturation value by using the notion of Hubbard projective measurement in the quantum teleportation protocol.

\subsection{Teleportation protocol with Hubbard projective measurement }
\begin{figure}[t!]
  \centering
  \subfigure[]{\includegraphics[width=0.5\textwidth]{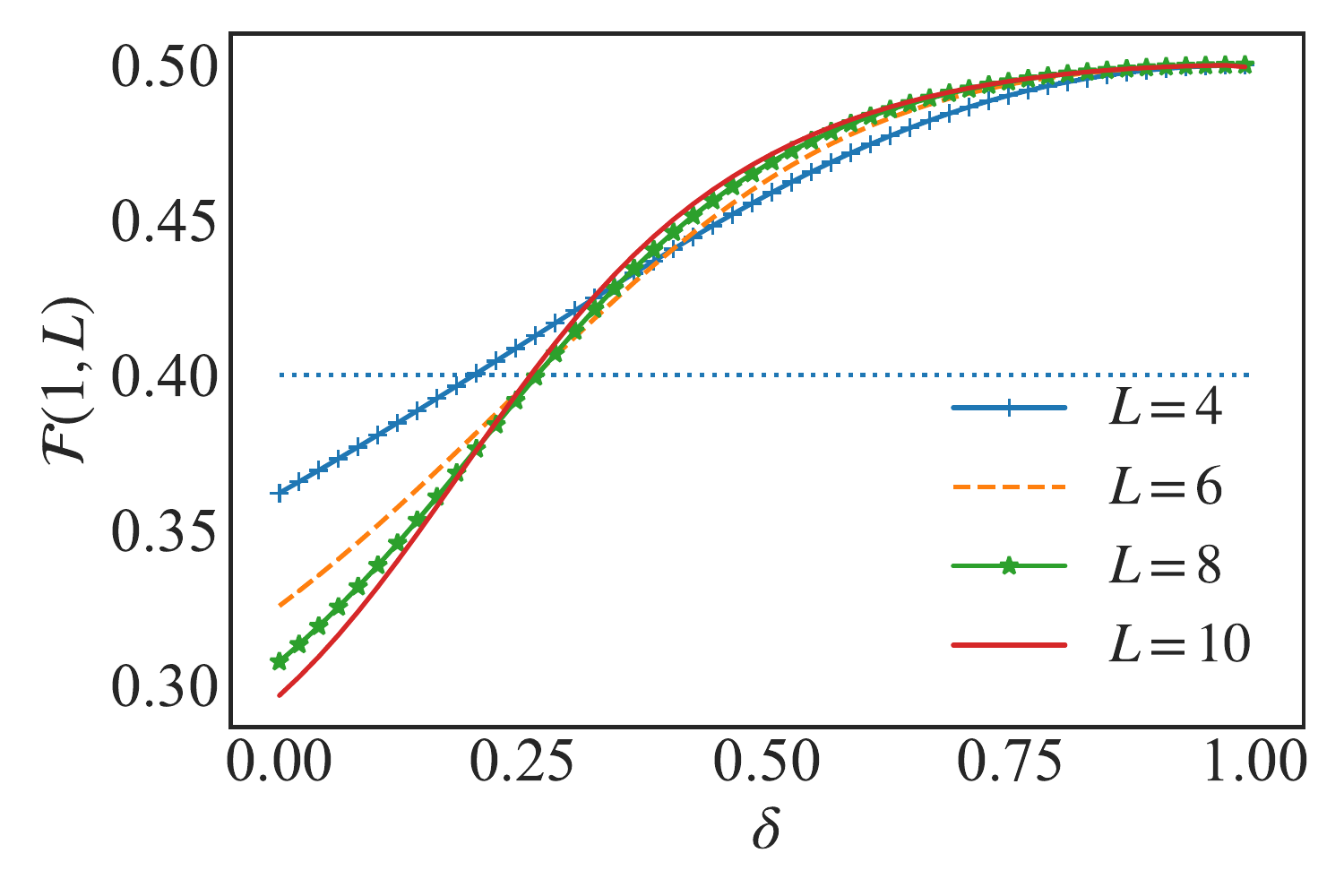}
  \label{fig3a}}%
  \subfigure[]{\includegraphics[width=0.5\textwidth]{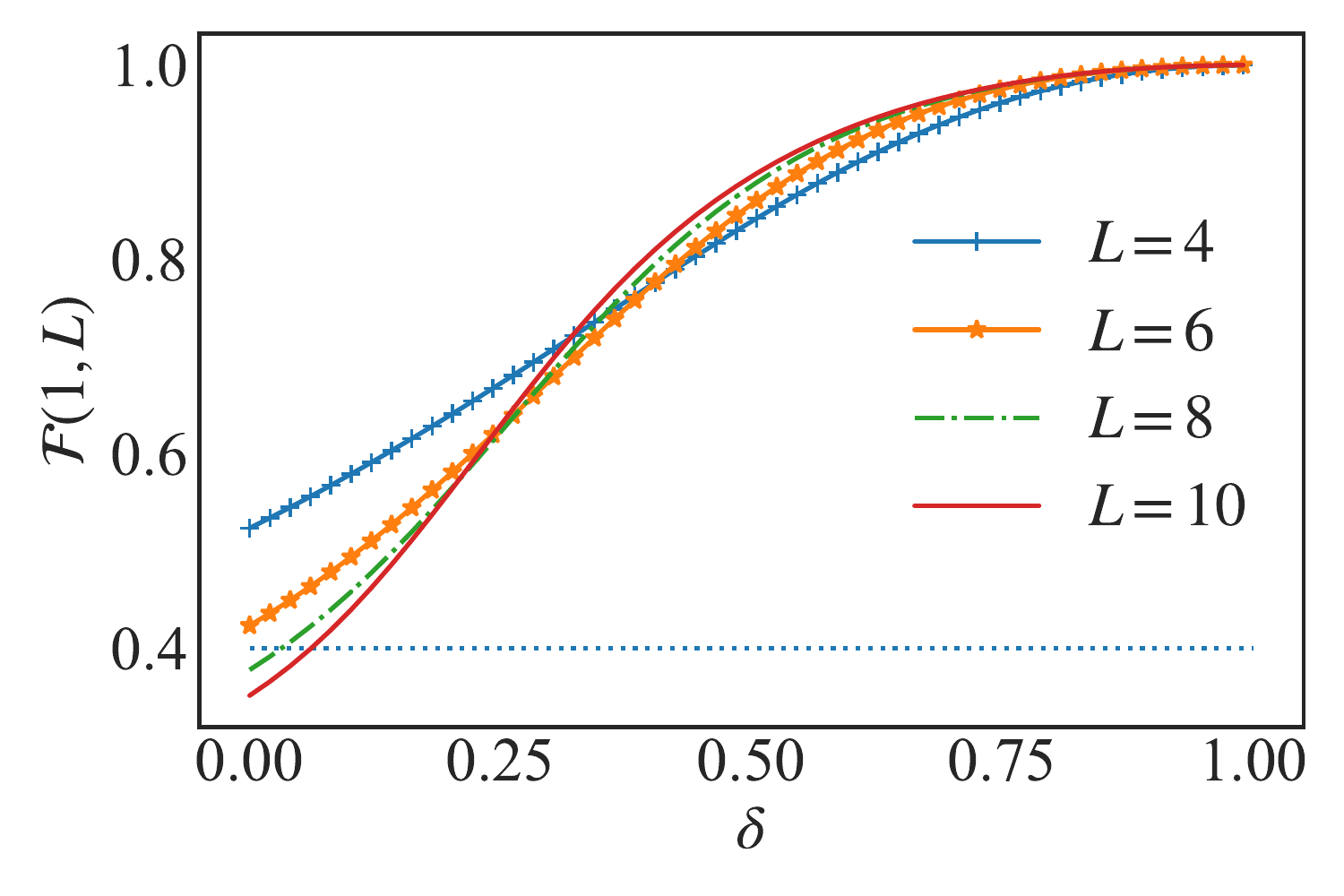}
  \label{fig3b}}
  \caption{The fidelity, Eq.~\eqref{fidelity}, with respect to $\delta$ for several chain sizes. In (a) Bell state projection are performed, while (b) consists of Hubbard projective measurements   }
  \label{fig3}
\end{figure}
We have demonstrated that the ground state of the Fermi-Hubbard, Eq.~\eqref{ham_delta}, with alternating weak and strong bonds allows for efficient quantum teleportation with a fidelity showing quantum advantage. The results found overlap with the data generated for creating long distance entanglement in the ground state of the Heisenberg spin chain, by alternating the interaction $J$ between the sites~\cite{Campos_Venuti}.
However, for the Fermi-Hubbard model a maximally entangled state with concurrence $C(\rho_{1L})\!=\!1$ seems to be insufficient for teleportation with unit fidelity and only $50\%$ of the information is transferred. Indeed, such a behavior is related to the inappropriate basis choice in the measurement protocol. Notably, the maximally entangled Bell state projectors are not suitable for the Fermi-Hubbard states. This can be clearly seen from the the fully entangled fraction $f$, Eq.~\eqref{fraction}, where $f\!=\!1$ only if the state $\ket{\psi^+}$ and the state of the entangled resource are identical. Therefore, we propose an adequate projective measurements based on the maximally entangled Hubbard state, i.e. the local half filled state. This allows for the generation of a new framework based on the inherent properties of the Hubbard model which is more convenient for teleportation with unit fidelity. This is confirmed in Fig.~\ref{fig3b}, where the end-to-end fidelity rise above the classical threshold (dotted line) quickly and saturates at the unit value as $\delta\!\to\!1$.

\subsection{Hubbard Model with alternating hopping amplitudes}

\begin{figure}[b!]
    \centering
    \includegraphics[trim={0cm 5cm 0cm 5cm}, clip, width=0.35\textwidth ]{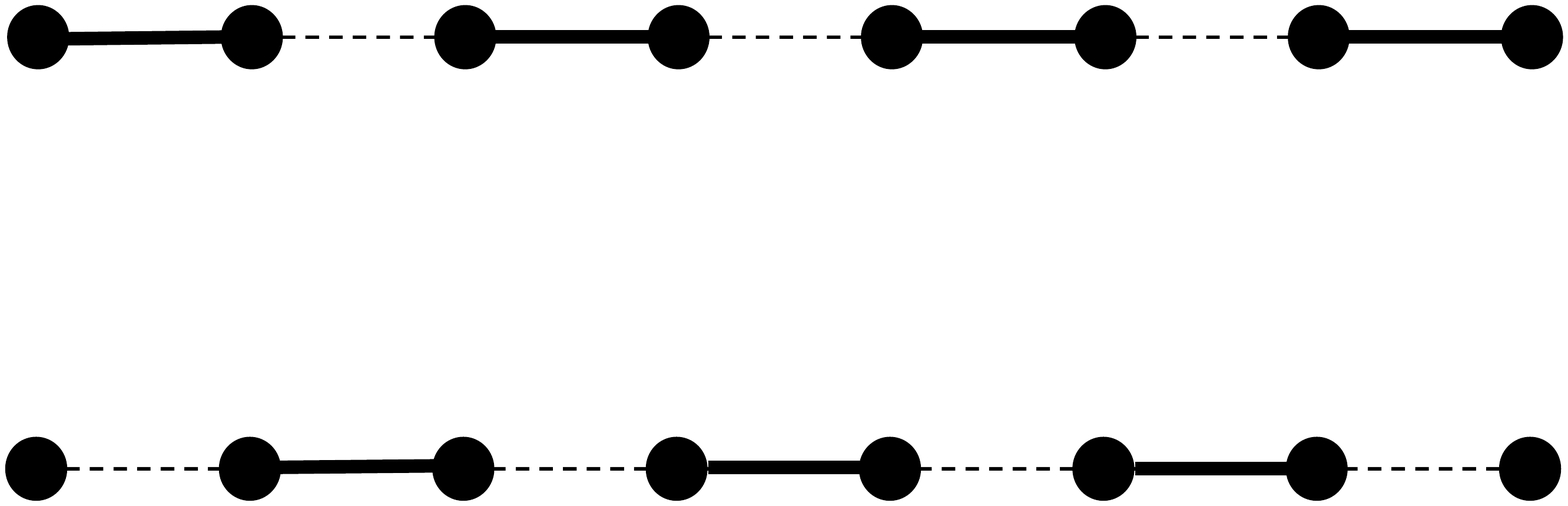}
    \put(-3,59){$\tau_a > \tau_b$}
    \put(-3,13){$\tau_a < \tau_b$}
    \put(-160,49){$\tau_a$}
    \put(-115,49){$\tau_a$}
    \put(-73,49){$\tau_a$}
    \put(-27,49){$\tau_a$}
    \put(-140,69){$\tau_b$}
    \put(-95,69){$\tau_b$}
    \put(-51,69){$\tau_b$}

    \caption{Sketch of the hopping bonds in the Hubbard chain, Eq.~\eqref{hamtatb}, with non uniform alternating $\tau_{a}$ and $\tau_{b}$.}
    \label{fig:img3}
\end{figure}
\begin{figure*}[t!]
  \centering
  \subfigure[]{\includegraphics[width=0.5\textwidth]{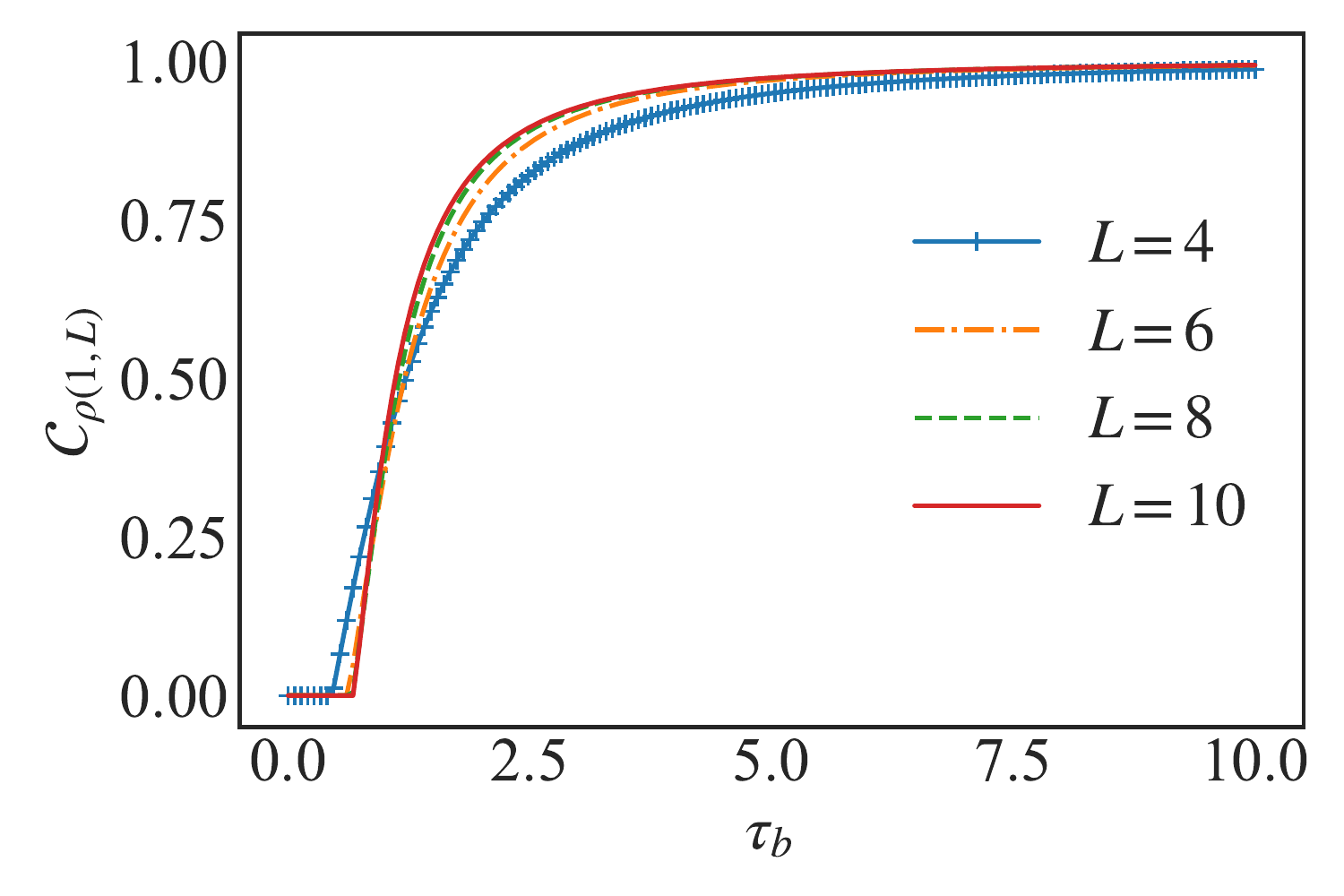}
  \label{fig:img4a}}%
  \subfigure[]{\includegraphics[width=0.5\textwidth]{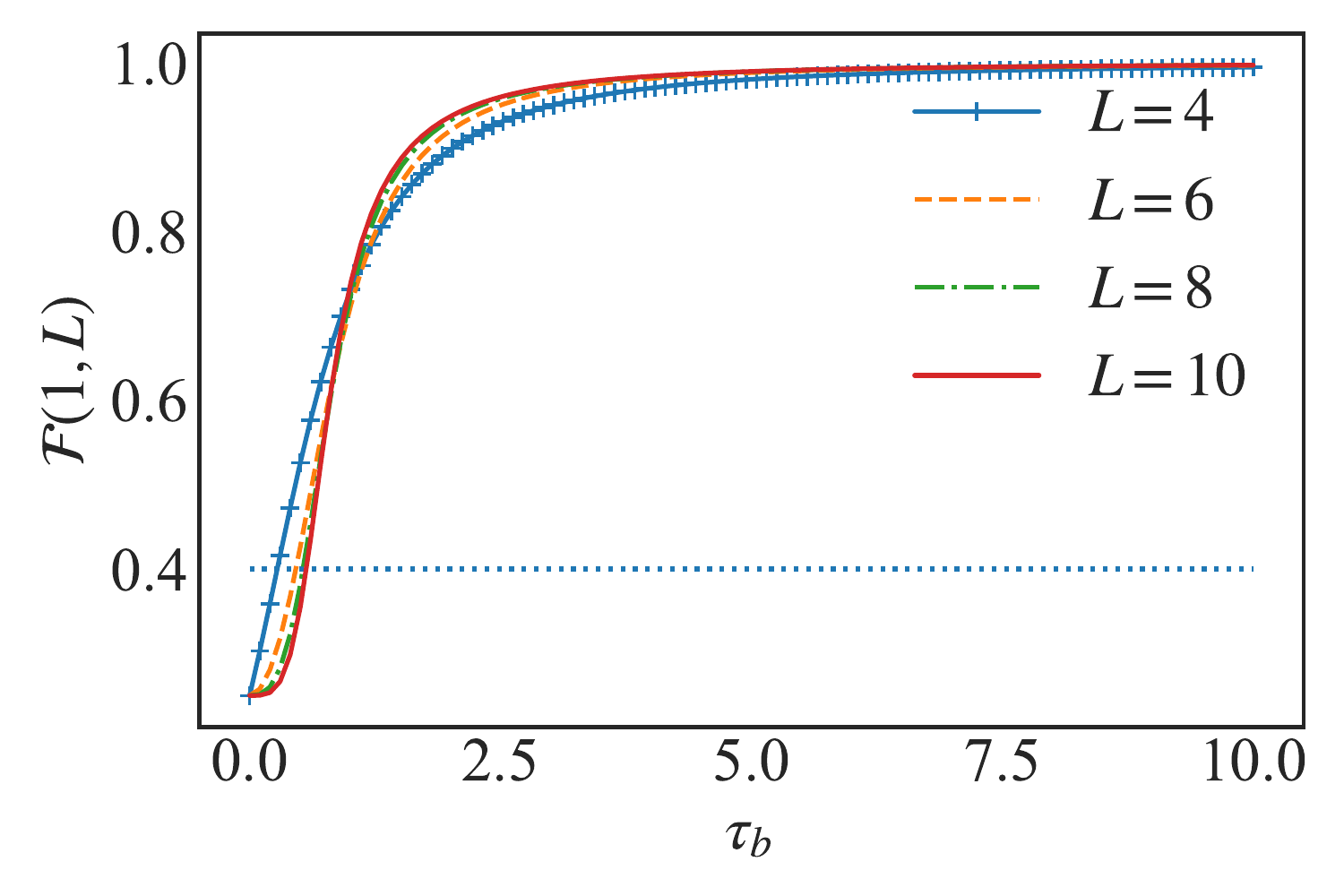}
  \label{fig:img4b}}\\
  \subfigure[]{\includegraphics[width=0.5\textwidth]{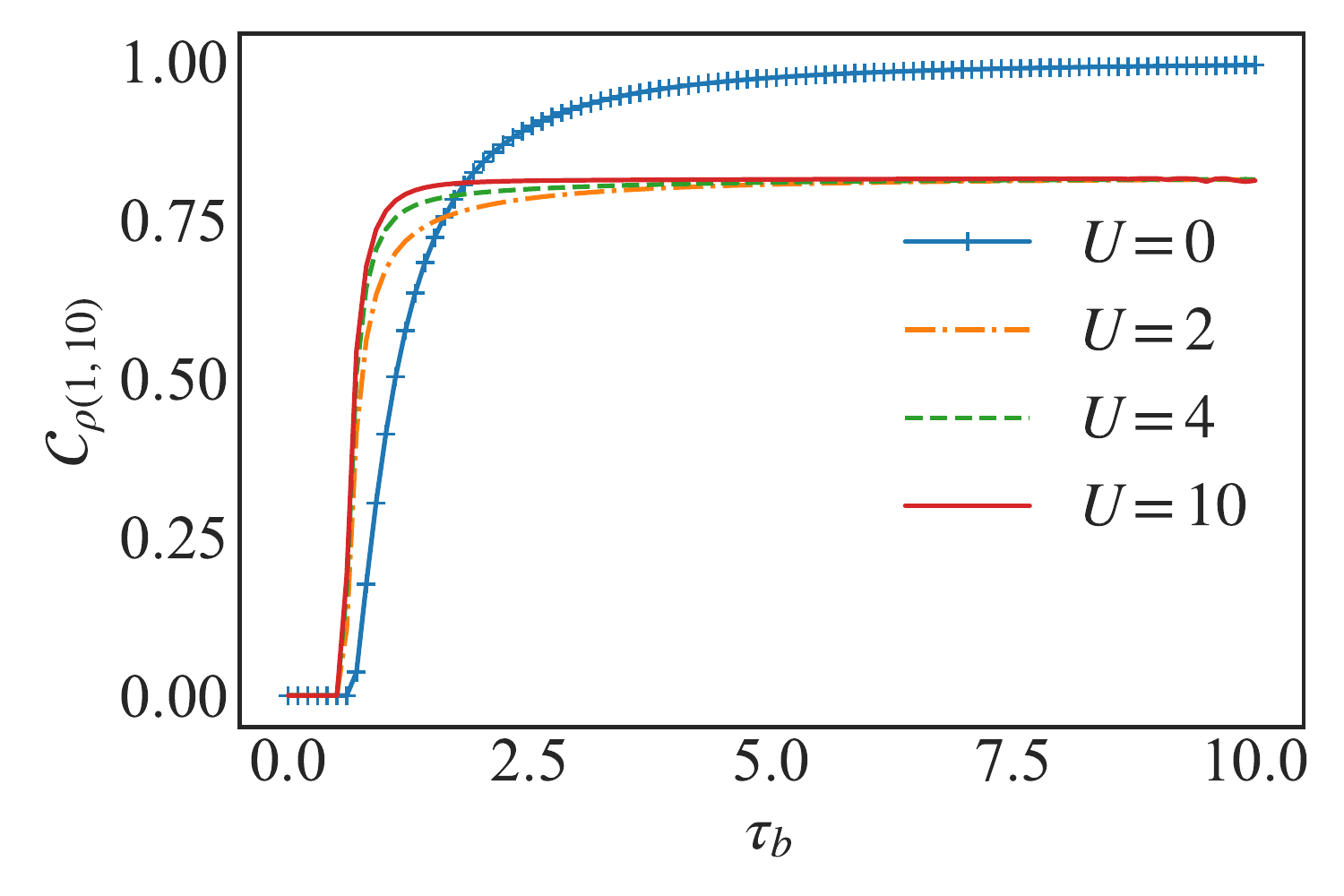}
  \label{fig:img4c}}%
  \subfigure[]{\includegraphics[width=0.5\textwidth]{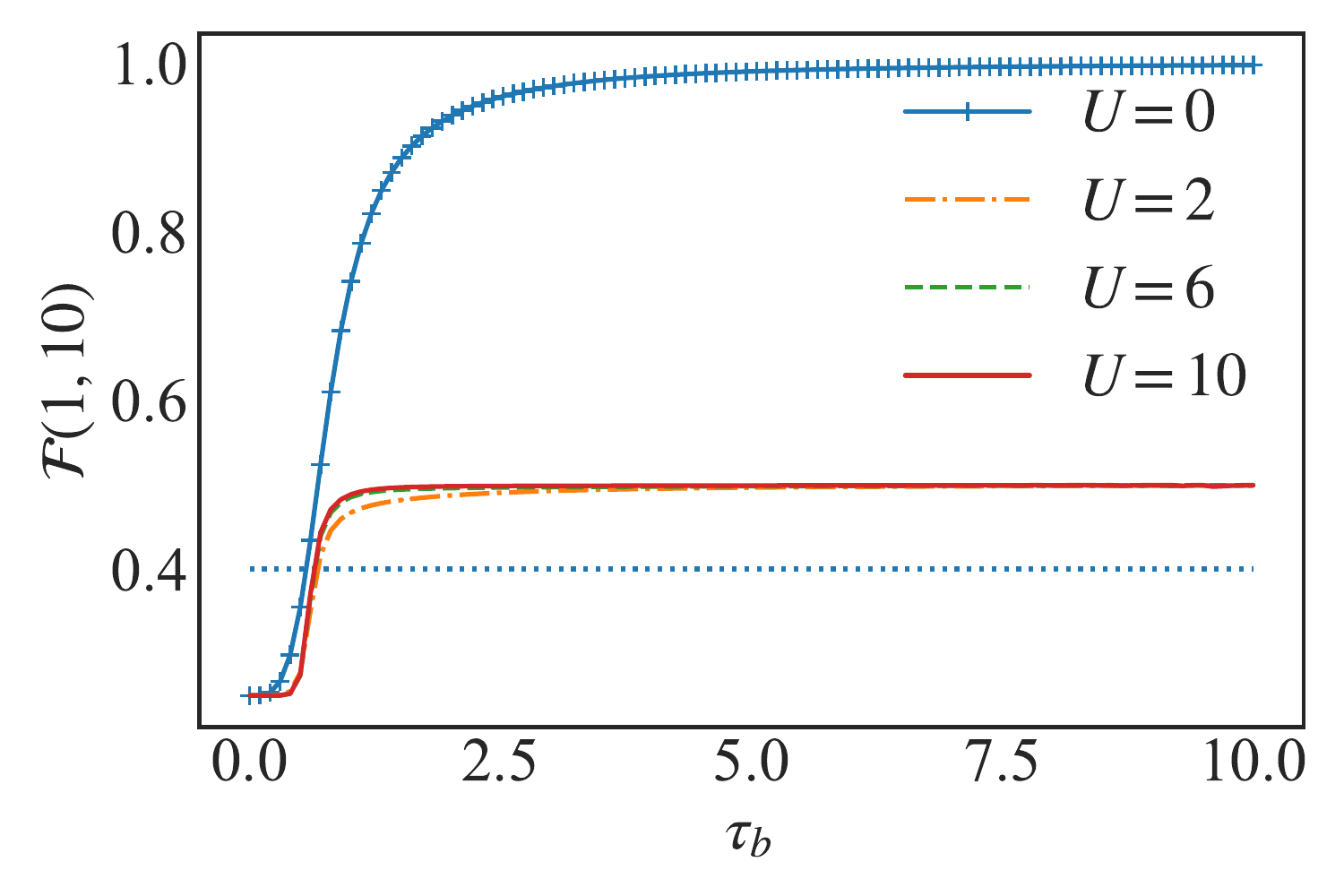}
  \label{fig:img4d}}
  \caption{(a)+(b) The end-to-end concurrence, Eq.~\eqref{lbc}, and fidelity, Eq.~\eqref{fidelity}, respectively, with respect to $\tau_b$ for several chain sizes $L$. (c)+(d) The same functions for different values of $U$ and fixed chain length $L$. }
  \label{fig:img4}
\end{figure*}

In this part, we discuss another way to create long distance entanglement by modeling the hopping interaction of the Fermi-Hubbard model via alternating hopping amplitudes $\tau_a$ and $\tau_b$, as sketched in Fig.~\ref{fig:img3}. In this case the Hamiltonian can be written as:

\begin{equation}
  H = - \sum_{i,\sigma} \tau_{a} \left( c_{2i-1,\sigma}^{\dag} c_{2i,\sigma}+ c_{2i,\sigma}^{\dag} c_{2i-1,\sigma} \right) +
  \tau_{b} \left( c_{2i,\sigma}^{\dag} c_{2i+1,\sigma}+ c_{2i+1,\sigma}^{\dag} c_{2i,\sigma} \right) + u \sum_{i} n_ {i,\uparrow} n_{i,\downarrow}.    
   \label{hamtatb}
\end{equation}
Fig.~\ref{fig:img4a} shows that when $ \tau_{b}\!<\!\tau_{a}$ no end-to-end entanglement is produced and this is related to the fact that the state of the pairs at borders $\rho_{1,2}$ and $\rho_{L-1,L}$ in this case are strongly correlated and there is no entanglement to share between the end sites. Increasing $ \tau_{b}$, the end-to-end entanglement grows rapidly starting from a threshold value that satisfies $\tau_{b_{T}}\ge \tau_{a}$, and is dependent on the size of the chain $L$. This behaviour translates in the fidelity of teleportation which grows more rapidly to 1 while increasing $\tau_b$ for smaller chain sizes $L$, as shown in Fig.~\ref{fig:img4b}. The effect of the Coulomb interaction on the end-to-end entanglement for a chain of $L=10$ is depicted in Fig.~\ref{fig:img4c}, where we see the threshold value is independent of $U$ when $U\!>\!0$. Additionally, the end-to-end concurrence rise rapidly to 1 for zero Coulomb interaction, while it reaches the asymptotic value of $0.8$ for $U\!>\!0$. Here again, the fidelity of teleportation follow the behavior of entanglement shown in Fig.~\ref{fig:img4d} for $L=10$, where we see that the slight reduction in the amount of the entanglement between the ends of the chain due to the Coulomb interaction, translates into large losses in the fidelity of the quantum teleportation. We note here that the Hubbard projective measurements have been used in the protocol in order to achieve unit fidelity of the quantum teleportation, which in combination with the alternating hopping amplitudes renders the protocol independent of the input state to be teleported (c.f. Figure 2 in the supplementary materials). 

\section{\label{sec: section4} Conclusion}
In summary, we have examined the ability of the one-dimensional Fermi-Hubbard model to support long distance entanglement in order to exploit its ground state as a channel for quantum teleportation between distant parties. To achieve the goal, we considered the Fermi-Hubbard model with bonds of alternating strengths, and alternating hopping amplitudes. We have established, for both cases, that long distance entanglement can hold with a maximum unit value independently of the system size, only for zero Couloumb interaction. Exploiting the property of long distance entanglement generation in the ground state of the Fermi-Hubbard model, and considering the fact that the quantum states of such model are four dimensional states, we have successfully demonstrated the capability of the one dimensional Fermi-Hubbard chains to operate as quantum channels for four dimensional state teleportation. Finally, we showed the crucial role of the measurement basis in the standard teleportation protocol, where a unit fidelity cannot be attained only by choosing an adequate basis choice based on the inherent properties of the chosen quantum channel, i.e maximally entangled Hubbard states. Our results motivate the investigation of long distance entanglement in the ground state of the Bose-Hubbard model and the Fermi-Bose-Hubbard model in order to inspect the role of symmetry of the ground state on the generation of entanglement, and on the fidelity of information transfer.
\section{Acknowledgements}
S.A. acknowledges gratefully the National Center for Scientific and Technical Research (CNRST) for financial support (Grant No.~1UM5R2018). Z.M. acknowledges support from the National Science Center (NCN), Poland, under Project No.~2020/38/E/ST3/00269. This research is supported through computational resources of HPC-MARWAN~\href{www.marwan.ma/hpc}{(www.marwan.ma/hpc)} provided by CNRST, Rabat, Morocco. 

\bibliography{biblio}
\section{Author contributions statement}
The authors confirm contribution to the paper as follows: study conception and
data collection: S.A, Z.M; analysis and
interpretation of results: S.A, Z.M, M.E.B; draft manuscript
preparation: S.A, Z.M, M.E.B. All authors reviewed the results and approved
the final version of the manuscript.
\section{Additional information}
The authors declare no competing interests.
\end{document}


\maketitle
\section{Standard Teleportation Protocol}

\begin{figure}[h]
 \centering
    \includegraphics[width=0.5\textwidth]{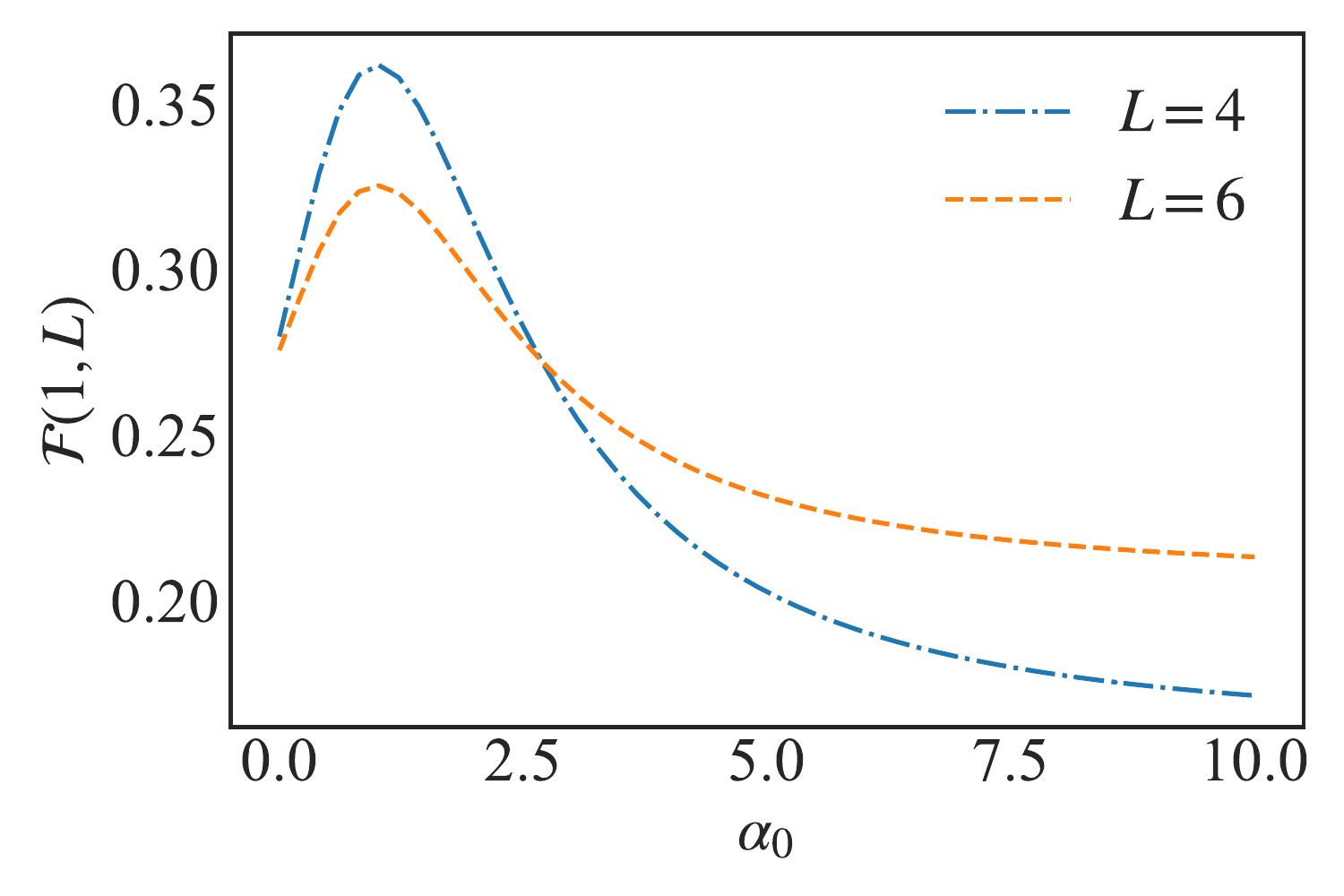}
    \caption{The end-to-end fidelity of the standard quantum teleportation channel as a function of the coefficient $\alpha_0$ for various system's sizes $L$ and $U=0$, using Bell state projective measurements.}
    \label{fig:fig0}
\end{figure}

\section{Teleportation protocol with Hubbard Projective Measurements}
\begin{figure}[h]
        \subfigure[\label{fig2a}]{%
       \includegraphics[width=0.33\textwidth]{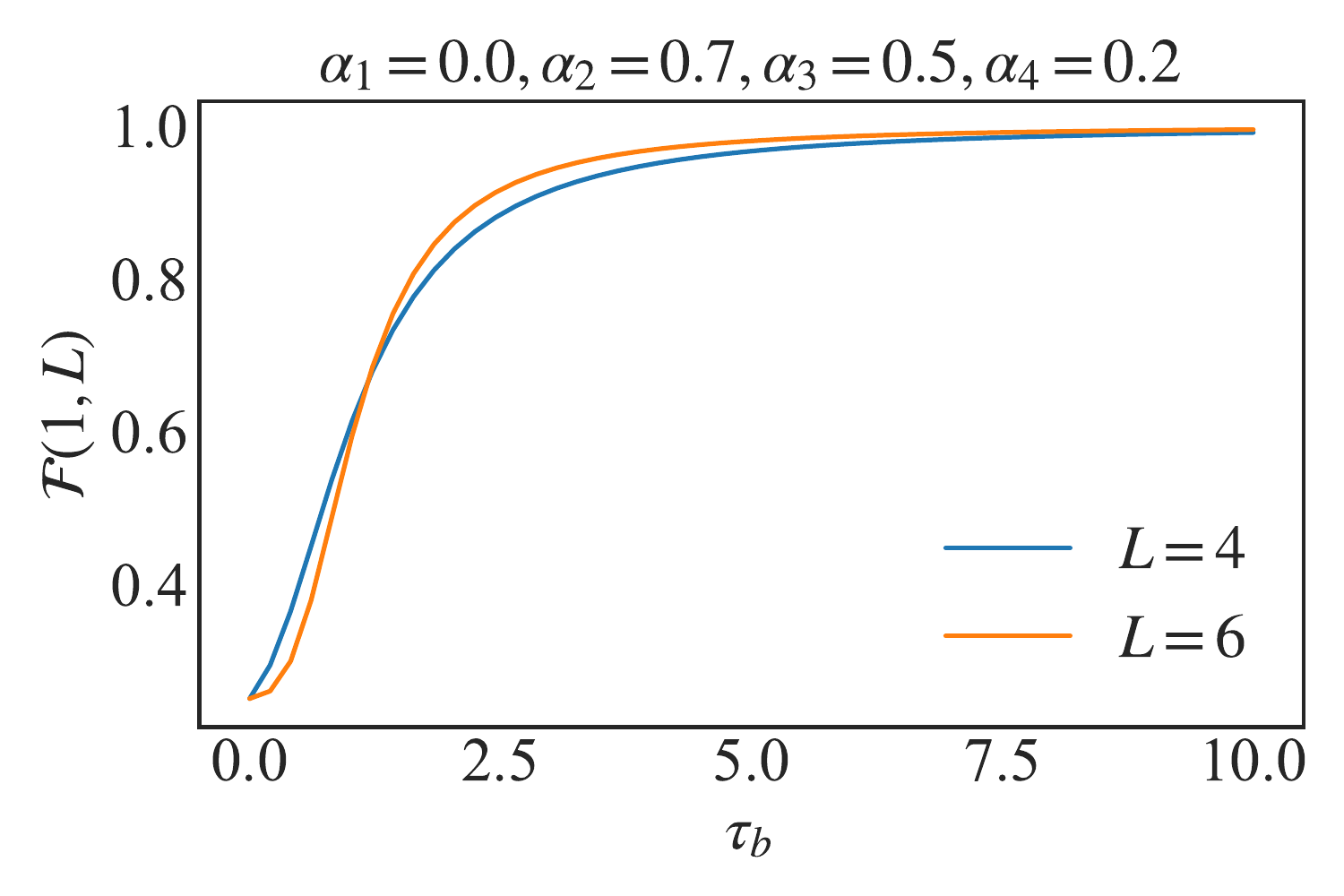}
       }%
     \subfigure[\label{fig2b}]{%
       \includegraphics[width=0.33\textwidth]{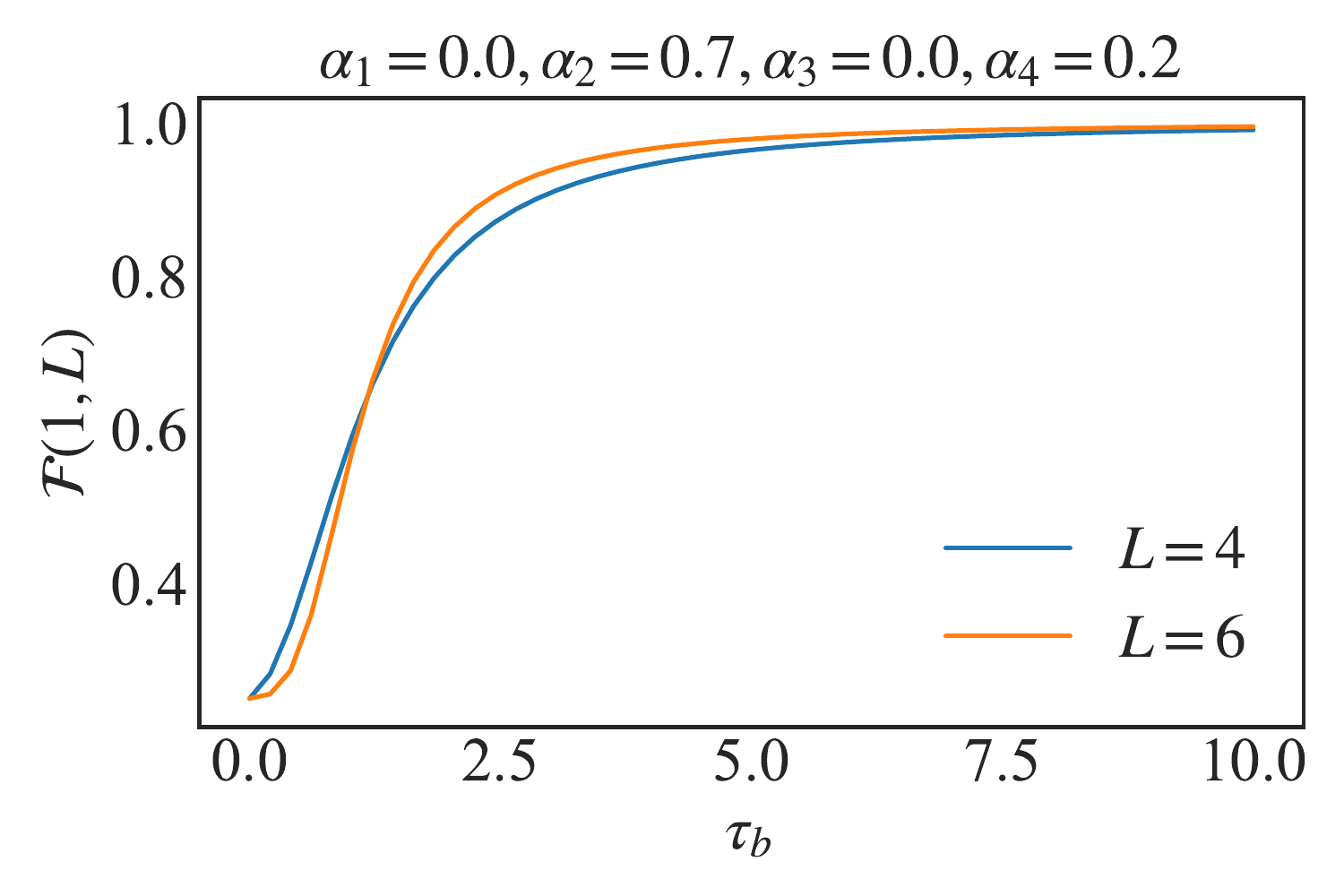}
       }%
     \subfigure[\label{fig2c}]{%
       \includegraphics[width=0.33\textwidth]{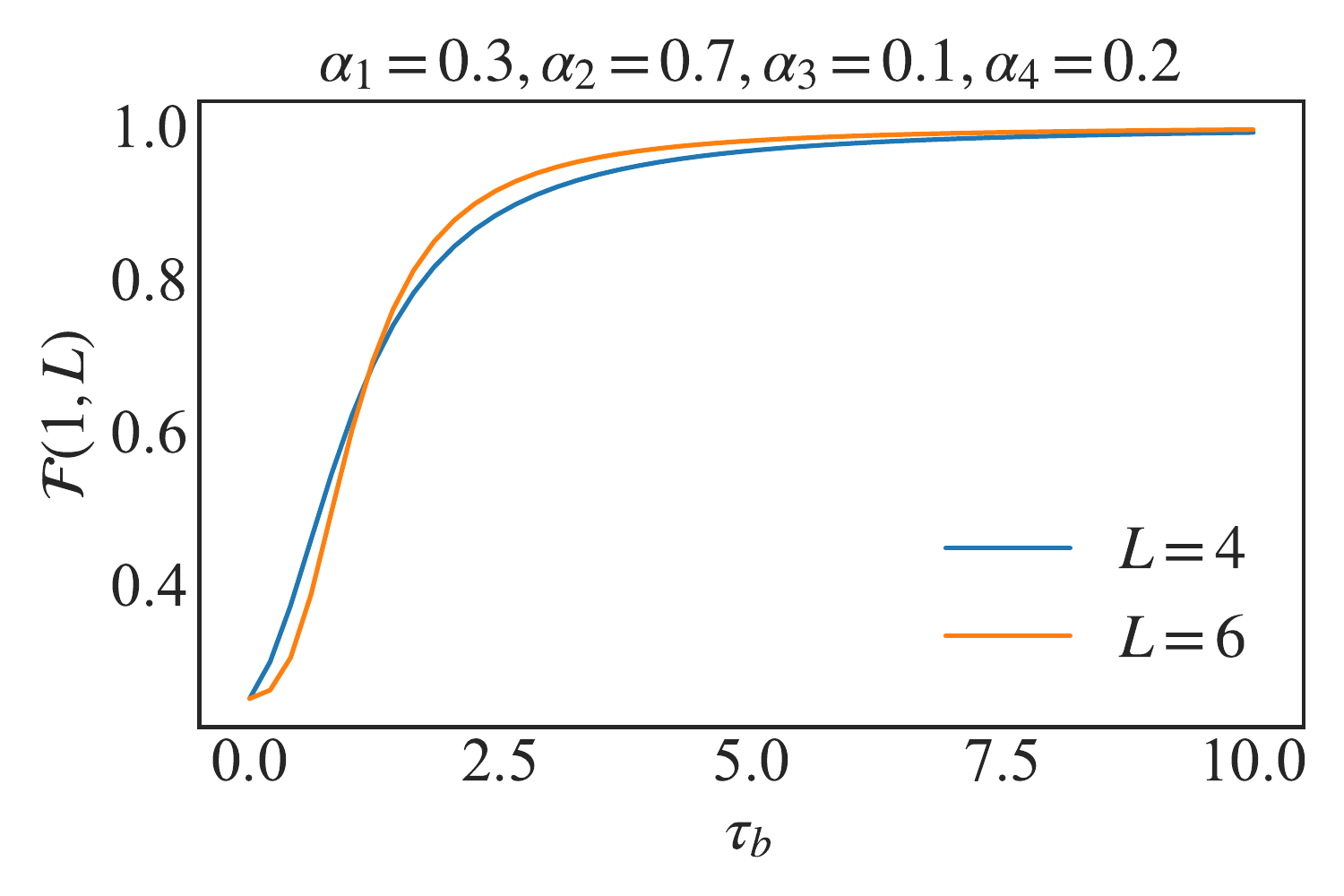}
       }
     \caption{The end-to-end fidelity of the standard quantum teleportation protocol using Hubbard projective measurements, for different values of the coefficients $\alpha_i$, with $i=0,1,2,3$ and various sizes $L$. }
     \label{fig2}
\end{figure}